\documentclass{PoS}

\usepackage{amsmath, graphics, graphicx}
\usepackage{natbib}
\usepackage{caption}
\usepackage{subcaption}
\usepackage{floatrow}
\title{Limits of noise and confusion in the MWA GLEAM year 1 survey}

\ShortTitle{Limits of noise and confusion in the MWA GLEAM year 1 survey}

\author{\speaker{Thomas~M.~O.~Franzen},$^1$ Carole~A.~Jackson,$^1$ Joseph~R.~Callingham,$^{3,2,4}$
Ron~D.~Ekers,$^{1,2}$ Paul~J.~Hancock,$^{1,2}$ Natasha~Hurley-Walker,$^1$ John~Morgan,$^1$
Nick~Seymour,$^1$ Randall~B.~Wayth,$^{1,2}$ Sarah~V.~White,$^{1}$ Martin~E.~Bell,$^{4,2}$ K.~S.~Dwarakanath,$^5$
Bi-Qing~For,$^6$ Bryan~M.~Gaensler,$^{7,2}$ Luke~Hindson,$^8$ Melanie~Johnston-Hollitt,$^8$
Anna~D.~Kapinska,$^{6,2}$ Emil~Lenc,$^{3,2}$ Ben~McKinley,$^{9,2}$ Andr\'{e}~R.~Offringa,$^{10}$
Pietro~Procopio,$^{9,2}$ Lister~Staveley-Smith,$^{6,2}$ Chen~Wu,$^6$ and Qian~Zheng$^8$\\
\llap{$^1$}ICRAR Curtin University, Australia;~ 
\llap{$^2$}ARC Centre of Excellence for All-sky Astrophysics (CAASTRO);~
\llap{$^3$}University of Sydney, Australia;~
\llap{$^4$}CSIRO Astronomy and Space Science (CASS), Australia;~
\llap{$^5$}Raman Research Institute, India;~
\llap{$^6$}ICRAR University of Western Australia, Australia;~
\llap{$^7$}Dunlap Institute for Astronomy \& Astrophysics, University of Toronto, Canada;~
\llap{$^8$}Victoria University of Wellington, New Zealand;~
\llap{$^9$}The University of Melbourne, Australia;~
\llap{$^{10}$}Netherlands Institute for Radio Astronomy (ASTRON), The Netherlands\\
E-mail: \email{thomas.franzen@curtin.edu.au}
}

\abstract{The GaLactic and Extragalactic All-sky MWA survey (GLEAM) is a new relatively low resolution, contiguous 
72--231~MHz survey of the entire sky south of declination $+ 25^{\circ}$. In this paper, we outline one approach 
to determine the relative contribution of system noise, classical confusion and sidelobe confusion in GLEAM images.
An understanding of the noise and confusion properties of GLEAM is essential if we are to fully exploit GLEAM data and improve
the design of future low-frequency surveys. Our early 
results indicate that sidelobe confusion dominates over the entire frequency range, implying 
that enhancements in data processing have the potential to further reduce the noise.}

\FullConference{EXTRA-RADSUR2015 (*)\\
		20--23 October 2015\\
		Bologna, Italy

                \bigskip
                \hrule
                \bigskip

                \textnormal{(*) This conference has been organized
                  with the support of the Ministry of Foreign Affairs
                  and International Cooperation, Directorate General
                  for the Country Promotion (Bilateral Grant Agreement
                  ZA14GR02 - Mapping the Universe on the Pathway to
                  SKA)}
}

\begin{document}

\section{Introduction}\label{Introduction}

The Galactic and Extragalactic All-sky MWA survey \citep[GLEAM;][]{wayth2015}, conducted with the Murchison Widefield Array 
\citep[MWA;][]{tingay2013}, covers the declination range $- 80^{\circ}$ to $+ 25^{\circ}$ at 72--231~MHz.
GLEAM observing began in August 2013 and its primary output from the first year of observations is
a catalogue of approximately 300,000 extragalactic radio components (Hurley-Walker et al., in preparation).

Large-area ($> 100~\mathrm{deg}^{2}$) surveys at low frequencies ($\lesssim 200$~MHz) 
such as GLEAM are limited by confusion effects at the mJy level, 
mainly due to large instrumental beam sizes. The situation is expected to improve with the extensive baselines and sensitivity 
of the Low Frequency Array \citep[LOFAR;][]{van_haarlem2013} and Square Kilometre Array Low \citep{dewdney2012}, which should 
push this limit substantially fainter.

There are three basic sources of error in a low-frequency image formed with an array: the system noise, classical 
confusion and sidelobe confusion, where we take sidelobe confusion to include calibration errors. 
In this paper, we analyse the relative contribution of system noise, classical confusion
and sidelobe confusion as a function of frequency in one of the deepest regions of GLEAM. 
An understanding of the noise and confusion properties is essential if we are to fully exploit GLEAM for extragalactic
radio source studies. This is also important for assessing whether enhancements in the data processing, 
such as improved deconvolution techniques, have the potential to further reduce the noise.

\section{MWA GLEAM observations and imaging strategy}\label{Observations and imaging strategy}

The MWA consists of 128 16-dipole antenna `tiles' distributed over an area approximately 3~km in diameter.
It operates at frequencies between 72 and 300~MHz, with an instantaneous bandwidth of 30.72~MHz.
Given the effective width ($\approx 4$~m) of the MWA's antenna tiles, the primary beam FWHM at 154~MHz is 27~deg.
The angular resolution, using a uniform image weighting scheme, is approximately $2.5 \times 2.2~\mathrm{sec}(\delta + 26.7^{\circ})$~arcmin. 
The excellent snapshot $uv$ coverage of the MWA and its huge field-of-view (FoV) allow it to rapidly image large areas of sky.

The survey strategy is outlined in \cite{wayth2015} while the data reduction and analysis methods
for the GLEAM year 1 catalogue will be described in detail in Hurley-Walker et al., in preparation. 
Briefly, meridian drift scans were used to cover the entire sky visible to the MWA.
The sky was divided into seven declination strips and five 30.72~MHz bands.
The observations were conducted as a series of 2-min scans for each frequency,
cycling through all five frequency settings over 10~min. The 2-min snapshots were divided into four 7.68~MHz
subbands and imaged separately with robust = --1 weighting using \textsc{WSClean} \citep{offringa2014}.
The snapshots for each subband were corrected for the primary beam and mosaicked together.

A deep wideband image covering $170-231$~MHz was formed by combining the eight highest frequency subband images.
This wideband image was used for source detection and flux density estimates were performed across the
20 7.68-MHz subbands. This approach maximised the number of sources catalogued and 
provided measurements for them across the full frequency range.

\section{Noise contribution}\label{Noise contribution}

In Section~\ref{System noise}, we follow \cite{wayth2015} to estimate the system noise in the GLEAM 7.68~MHz subband mosaics
in a `cold' region of extragalactic sky near zenith. Observations with the Giant Metrewave Radio Telescope
by \cite{intema2011}, \cite{ghosh2012} and \cite{williams2013} probe the 153~MHz counts down to 6, 12 and 15~mJy, 
respectively. In Section~\ref{Classical confusion}, we use these deep source counts to estimate the
classical confusion noise across the GLEAM frequency range.
Finally, in Section~\ref{Sidelobe confusion}, we compare our estimates of the system noise and classical confusion noise 
with the measured rms noise to draw conclusions about the degree of sidelobe confusion.

\subsection{System noise}\label{System noise}

The system noise, $T_{\mathrm{sys}}$, is the Gaussian random noise resulting from the noise power entering the system
and is equal to $T_{\mathrm{sky}} + T_{\mathrm{rec}}$, where $T_{\mathrm{sky}}$ is the sky noise and $T_{\mathrm{rec}}$ the receiver noise.
For the MWA, $T_{\mathrm{sys}}$ is dominated by $T_{\mathrm{sky}}$ over the entire frequency range covered by GLEAM,
with a much smaller contribution from $T_{\mathrm{rec}}$ ($\approx 50$~K). It follows that $T_{\mathrm{sys}}$
strongly depends on the region of sky being observed. 

\begin{figure}
\includegraphics[scale=0.3, trim=0cm 0cm 0cm 0cm, angle=270]{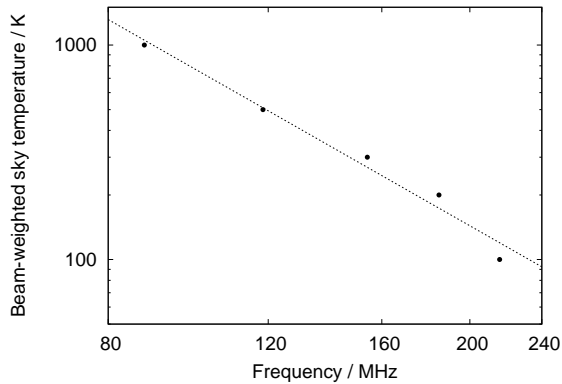}
\caption{Beam-weighted sky temperature as a function of frequency in a `cold' region of extragalactic sky at $\delta = -26.7$~deg
and LST = 0~h. The dashed line is a power-law fit to the data points.}
\label{fig:tsky}
\end{figure}

The expected beam-weighted average sky temperature over the range of frequencies,
pointings and LSTs relevant to GLEAM was calculated by \cite{wayth2015}. Fig.~\ref{fig:tsky} shows the beam-weighted average sky temperature at 
the central frequency of each 30.72~MHz band in a `cold' region of extragalactic sky. We fit a power law to
these data points, obtaining a slope of --2.42.

From noise-only simulations that match the GLEAM observing strategy, \cite{wayth2015} estimated the expected
thermal noise for GLEAM 30.72~MHz band mosaics for a single fiducial system temperature $T_{\mathrm{f}}$. Table~\ref{tab:thermal_noise}
shows their thermal noise estimates for a pointing at declination --26.7~deg assuming $T_{\mathrm{f}} = 200$~K.
Our thermal noise estimates for GLEAM 7.68~MHz subband mosaics are shown in Fig.~\ref{fig:noise_vs_freq}; they were
obtained by multiplying the values in Table~\ref{tab:thermal_noise} by 
$\sqrt{\frac{30.72~\mathrm{MHz}}{7.68~\mathrm{MHz}}} \frac{T_{\mathrm{sys}} + T_{\mathrm{rec}}}{T_{\mathrm{f}}}
= 2 \frac{T_{\mathrm{sys}} + T_{\mathrm{rec}}}{T_{\mathrm{f}}}$, where $T_{\mathrm{sys}}$ is the system temperature
at the central frequency of the subband.

\begin{table}
\centering
\begin{tabular}{@{} c c} 
\hline
Frequency &
\multicolumn{1}{c}{Thermal noise sensitivity} \\
 (MHz) &
\multicolumn{1}{c}{(mJy/beam)} \\
\hline
87.7 & 1.7 \\
118.4 & 1.9 \\
154.2 & 2.1 \\
185.0 & 2.2 \\
215.7 & 2.4 \\
\hline
\end{tabular}
\caption{GLEAM expected thermal noise sensitivity at a declination of --26.7~deg, assuming a bandwidth of 30.72~MHz and $T_{\mathrm{f}} = 200$~K,
 as indicated by \cite{wayth2015}.}
\label{tab:thermal_noise}
\end{table}

\subsection{Classical confusion}\label{Classical confusion}

When the density of faint extragalactic sources becomes too high for them to be clearly resolved by the array, the
deflections in the image will include the sum of all the unresolved sources in the main lobe of the synthesised beam. 
This effect is known as classical confusion and only depends on the source counts and the synthesised beam area \citep{condon1974}. 

We use the method of probability of deflection, or $P(D)$ analysis \citep{scheuer1957}, to quantify the classical confusion,
$\sigma_{\mathrm{c}}$, in GLEAM.
The deflection $D$ at any pixel in the image is the intensity in units of mJy/beam. Given a source count model and 
synthesised beam size, we use Monte Carlo simulations to derive the exact shape of $P_{\mathrm{c}}(D)$, the $P(D)$ 
distribution from all sources present in the image, in each subband. We then estimate the rms classical confusion 
noise from the core width of this distribution.

\begin{figure}
\centering
\includegraphics[scale=0.4, trim=0cm 0cm 0cm 0cm, angle=270]{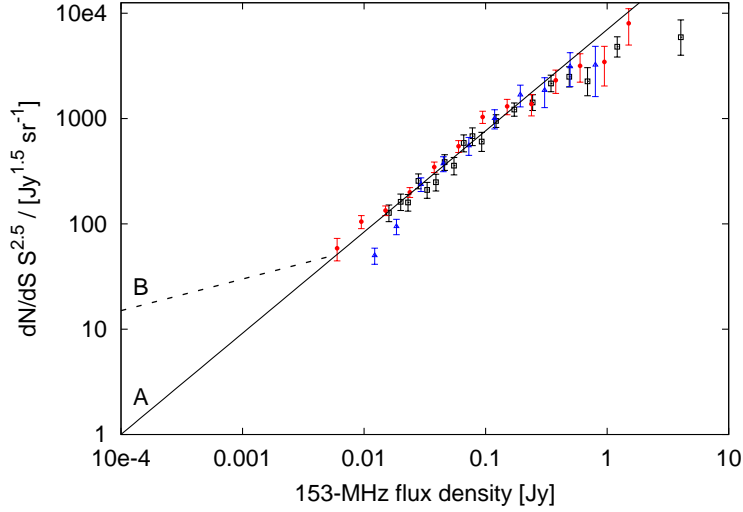}
\caption{The black squares, red circles and blue triangles show the Euclidean normalised differential counts 
at 153~MHz from \cite{williams2013}, \cite{intema2011} and \cite{ghosh2012}, respectively. 
The solid and dashed lines show source count models A and B, respectively.}
\label{fig:deep_150MHz_counts}
\end{figure}

In the flux density range 6--400~mJy, the Euclidean normalised differential counts at 153~MHz from \cite{williams2013},
\cite{intema2011} and \cite{ghosh2012} are well represented by a power law of the form $\frac{dN}{dS} = kS^{-\gamma} \, \mathrm{Jy}^{-1} \mathrm{sr}^{-1}$, 
with $k = 6998$ and $\gamma = 1.54$ (see Fig.~\ref{fig:deep_150MHz_counts}). The 
153~MHz differential source counts continue to decline at $S_{153} \lesssim 10$~mJy and no flattening of the differential count at low flux densities (e.g. as seen at 1.4~GHz) has been detected. We use two source count models in the
Monte Carlo simulations to derive $P_{\mathrm{c}}(D)$. The counts are modelled as
\begin{eqnarray}
\label{eqn:scount_model_formula}
n(S) \equiv \frac{\mathrm{d}N}{\mathrm{d}S} \approx
\left\{
\begin{array}{ll}
k_{1} \left(\frac{S}{\mathrm{Jy}}\right)^{-\gamma_{1}}~\mathrm{Jy^{-1}~sr^{-1}}~\mathrm{for}~S_{\mathrm{low}} \leq S < S_{\mathrm{mid}} \\
k_{2} \left(\frac{S}{\mathrm{Jy}}\right)^{-\gamma_{2}}~\mathrm{Jy^{-1}~sr^{-1}}~\mathrm{for}~S_{\mathrm{mid}} \leq S \leq S_{\mathrm{high}}~\mathrm{.}
\end{array}
\right.
\end{eqnarray}
The values of the source count parameters $k_{1}$, $\gamma_{1}$, $k_{2}$, $\gamma_{2}$, $S_{\mathrm{low}}$, $S_{\mathrm{mid}}$ 
and $S_{\mathrm{high}}$ for each model are provided in Table~\ref{tab:scount_model_par}. Model A corresponds to
the case where there is no flattening in the counts below 6~mJy. In model B, the source count slope is set to 2.2
below 6~mJy; the purpose of model B is to explore how sensitive the classical confusion noise is to a flattening
in the source count slope below 6~mJy.

\begin{table}
\centering 
\begin{tabular}{r r r r r r r r}
\hline
Model & $k_{1}$ & $\gamma_{1}$ & $k_{2}$ & $\gamma_{2}$ & $S_{\mathrm{low}}$ & $S_{\mathrm{mid}}$ & $S_{\mathrm{high}}$ \\
 &  &  &  &  & (mJy) & (mJy) & (mJy) \\
\hline
A & 6998 & 1.54 & 6998 & 1.54 & 0.1 & 6.0 & 400 \\
B & 237.9 & 2.200 & 6998 & 1.54 & 0.1 & 6.0 & 400 \\
\hline
\end{tabular}
\caption{Parameter values adopted to model the 153~MHz counts.}
\label{tab:scount_model_par}
\end{table}

We extrapolate the two 153~MHz source count models to the central frequency of each GLEAM subband, assuming each source varies 
with frequency as $S \propto \nu^{\alpha}$, by applying the method described in \cite{waldram2007}. We assume three different
values of $\alpha$: --0.5, --0.7 and --0.9.

\begin{figure}
\centering
\includegraphics[scale=0.35, trim=0cm 0cm 0cm 0cm, angle=270]{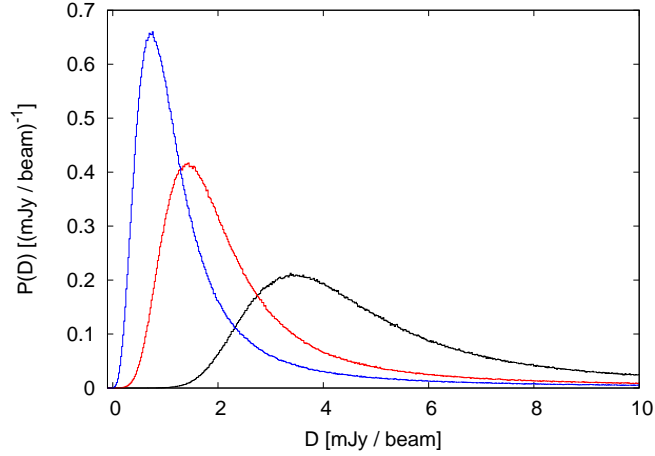}
\caption{$P_{\mathrm{c}}(D)$ distributions at 122~MHz (black), 158~MHz (red) and 189~MHz (blue) corresponding to source count model B and $\alpha = -0.7$.}
\label{fig:source_PofD_examples}
\end{figure}

We derive $P_{\mathrm{c}}(D)$ for each GLEAM subband and source count model as follows: we simulate a noise-free image 
containing point sources at random positions, assigning their flux densities according to the source count model.
Sources with flux densities ranging between 0.1 and 400~mJy are injected into the image using the \textsc{miriad} task \textsc{imgen} \citep{Sault1995}. The simulated point sources are convolved with the Gaussian restoring beam of the subband image; 
we do not attempt to model the sidelobe confusion. We obtain $P_{\mathrm{c}}(D)$ from the distribution of pixel values in the simulated image. The width of the distribution is measured by dividing the interquartile range by 1.349, i.e. the rms for a 
Gaussian distribution. Some examples of $P_{\mathrm{c}}(D)$ are shown in Fig.~\ref{fig:source_PofD_examples}.

\begin{figure}
\centering
\includegraphics[scale=0.35, trim=0cm 0cm 0cm 0cm, angle=270]{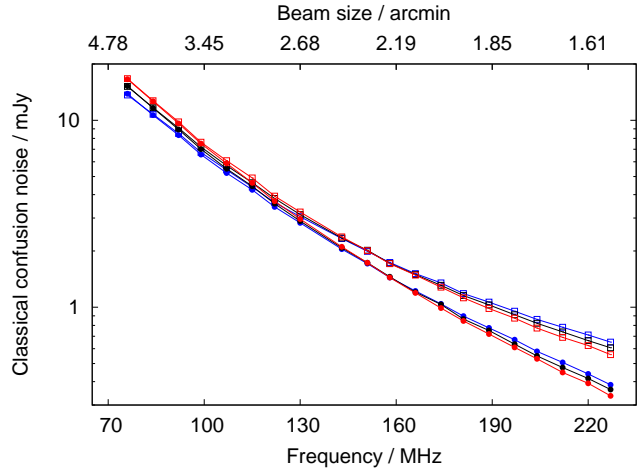}
\caption{The circles and squares show estimates of $\sigma_{\mathrm{c}}$ using source count models A and B, respectively.
These source count models are extrapolated to the GLEAM subband frequencies (72--231~MHz) assuming $\alpha = -0.5$ (blue),
$\alpha = -0.7$ (black) and $\alpha = -0.9$ (red).}
\label{fig:compare_classical_conf_noise}
\end{figure}

Fig.~\ref{fig:compare_classical_conf_noise} shows that models A and B diverge
at higher frequency, indicating that sources below 6~mJy are too faint to contribute 
significantly to the confusion noise except at the highest frequencies where the beam size is smallest.

\subsection{Sidelobe confusion}\label{Sidelobe confusion}

Additional noise is introduced into an image from the combined sidelobes of undeconvolved sources,
i.e. from the array response to sources below the source subtraction cut-off limit and to sources outside the imaged FoV.
This effect is known as sidelobe confusion.
Since the MWA has many baselines and no regularity in the aperture, 
sidelobes from any single short observation have a nearly random distribution in the image and are not easily 
distinguishable from other noise terms. In order to draw conclusions about the degree of sidelobe confusion, 
we compare our estimates of the system noise and classical confusion noise 
with the measured rms noise in one of the deepest regions of GLEAM.

\begin{figure}
\centering
\includegraphics[scale=0.5, trim=2cm 0cm 0cm 0cm, angle=270]{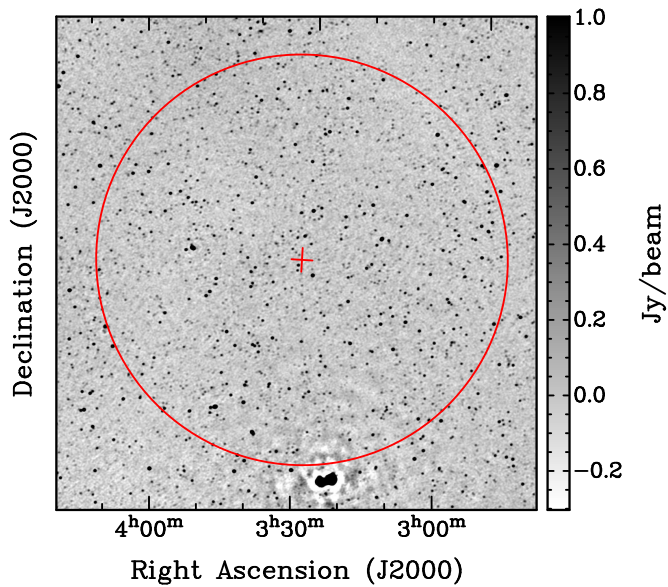}
\caption{Image of a section of the lowest GLEAM subband image at 72--80~MHz centred on CDFS. The region in which the mean rms noise is measured
is shown bounded in red.}
\label{fig:white_image}
\end{figure}

We measure the mean rms noise in a region within 8.5~deg of the \textit{Chandra} Deep Field-South (CDFS) 
at J2000 $\alpha = 03^{\mathrm{h}}30^{\mathrm{m}}00^{\mathrm{s}}$, $\delta = -28^{\circ}00'00''$.
This region lies close to zenith (i.e. at $\delta = -26.7$~deg) and 55~deg from the Galactic Plane.
As an example, Fig.~\ref{fig:white_image} shows the lowest GLEAM subband image of this region of sky. We use 
\textsc{BANE}\footnote{https://github.com/PaulHancock/Aegean} to calculate a noise image from the interquartile range
of pixels in regions of size $20 \times 20$ synthesised beams. 

\begin{figure}
\centering
\includegraphics[scale=0.45, trim=0cm 0cm 0cm 0cm, angle=270]{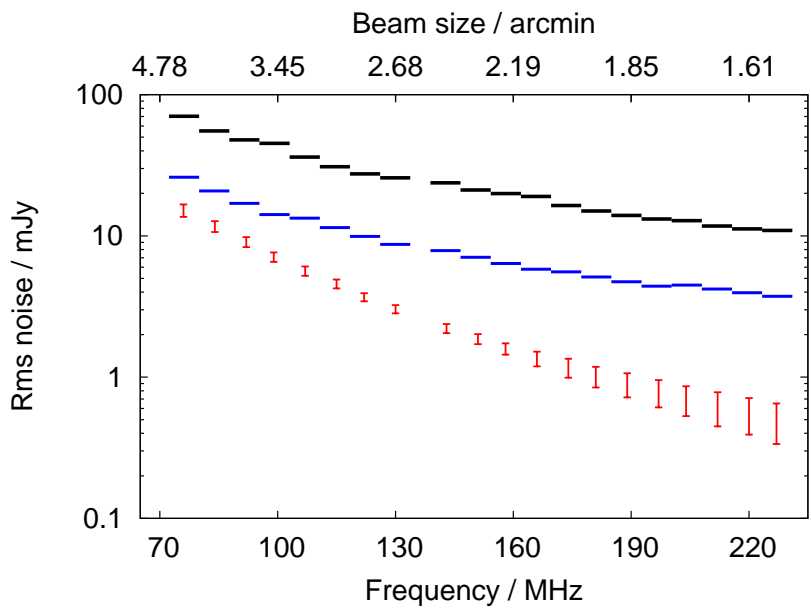}
\caption{Black horizontal bars: mean rms noise in the 7.68~MHz subband images
in a region within 8.5~deg from CDFS. The approximate beam size is
shown on the top. Blue horizontal bars: expected thermal noise sensitivity \citep{wayth2015}.
Red points: range of classical confusion noise estimates.}
\label{fig:noise_vs_freq}
\end{figure}

Fig.~\ref{fig:noise_vs_freq} reveals that the thermal and classical confusion noise are much 
lower than the measured noise at all frequencies. From this, we conclude that the background noise is primarily 
due to sidelobe confusion. Possible origins of the sidelobe confusion 
are the limited CLEANing depth, far-field sources that have not been deconvolved, and residuals of ionospheric smearing.

\section{Conclusions and future work}\label{Discussion}

Our initial work suggests that the background noise in GLEAM images is primarily due to sidelobe confusion. This is a consequence of the large
FoV and the huge number of detected sources. A similar result was found by \cite{franzen2016} in one of the deepest 
images ever made with the MWA, a single MWA pointing image of an Epoch of Reionisation field at 154~MHz with a resolution
of 2.3~arcmin \citep{offringa2016}. This image is affected by sidelobe confusion noise at the $\approx 3.5$~mJy/beam level, 
and the classical confusion limit is $\approx 1.7$~mJy/beam. 

In future work, we will determine the noise contribution in different regions of GLEAM (e.g. close to the Galactic plane, powerful
radio sources etc.), and investigate how these vary across 72--231~MHz and where best improvements may be made.

\acknowledgments

This scientific work makes use of the Murchison Radio-astronomy Observatory, operated by CSIRO. We acknowledge the Wajarri Yamatji people as the traditional owners of the Observatory site. CAJ thanks the Department of Science, Office of Premier \& Cabinet, WA 
for their support through the Western Australian Fellowship Program. Support for the operation of the MWA is provided by the Australian Government Department of Industry and Science and Department of Education (National Collaborative Research Infrastructure Strategy: NCRIS), under a contract to Curtin University administered by Astronomy Australia Limited. We acknowledge the iVEC Petabyte Data Store and the Initiative in Innovative Computing and the CUDA Center for Excellence sponsored by NVIDIA at Harvard University.

\bibliographystyle{apj}

\newcommand{\aapr}{Astron. Astrophys. Rev.}
\newcommand\nar{New Astronomy Reviews}
\newcommand{\nat}{Nat}
\newcommand{\sci}{Sci}

\setlength{\bibsep}{0.0pt}

\end{document}